\documentstyle[11pt,graphicx]{article}

\large

\title{ Relic gravitational waves
and their detection }
\author{\small Wen Zhao, Yang Zhang \\
        \small Astrophysics Center \\
        \small University of Science and Technology of China \\
        \small Hefei, Anhui, China }
 \date{}

\begin{document}
\maketitle
\baselineskip=19truept
\def\vek{\vec{k}}

\newcommand{\be}{\begin{equation}}
\newcommand{\ee}{\end{equation}}
\newcommand{\ba}{\begin{eqnarray}}
\newcommand{\ea}{\end{eqnarray}}
\renewcommand{\H}{{\cal H}}
\renewcommand{\L}{{\cal L}}
\def\mpl{m_{\mathrm{Pl}}}
\def\Mpl{M_{\mathrm{Pl}}}

\sf \small

\begin{center}
\Large  Abstract
\end{center}
\begin{quote}

{ As strong evidence for inflation, the relic gravitational waves
(RGW) have been extensively studied. Although, it has not been
detected, yet some constraints have been achieved by the
observations. Future experiments for the RGW detection are mainly
two kinds: the CMB experiments and the laser interferometers. In
this paper, we study  these current constraints and the detective
abilities of future experiments. We  calculate the strength of RGW
$\Omega_g(k)$ in two methods: the analytic method and the
numerical method by solving the inflationary flow equations. By
the first  method  we obtain a bound $\Omega_g<3.89\times10^{-16}$
at $\nu=0.1$Hz, where we have used
 the current constraints on  the scalar spectral
index,  the tensor-scalar ratio,
furthermore, we have taken into account of the redshift-suppression
effect, the accelerating expansion effect, the neutrino damping
effect on the RGW.
But the analytic expression of  $\Omega_g(k)$ depends on
 the specific inflationary models
and only applies for the waves with very low frequencies.
The numerical  method
is more precise for the high frequency waves and applies to any
single-field inflationary model.
It gives a bound
$\Omega_g<8.62\times10^{-14}$, which is independent of the
inflationary parameters, and applies to any single-field slow-roll
inflationary model.
After considering the current constraints on
the inflationary parameters,
this bound reduces down to $\Omega_g<2\times10^{-17}$.
These two methods
give the consistent conclusions: The current constraints on the
RGW from LIGO, big bang nucleosynthesis,  and pulsar timing
are too loose to give any stringent constraint for the single-field
inflationary models, and the constraint from WMAP are relatively
tighter.
The future laser interferometers are more effective for
detecting the RGW with  the smaller tensor-scalar ratio,
but the CMB experiments are
more effective for detecting the waves with the larger ratio.
These detection methods  are
complementary to each other for the  detections of RGW. }
\end{quote}


e-mail: wzhao7@mail.ustc.edu.cn

\baselineskip=17truept

~

\newpage
\small

\begin{center}
{\large 1. Introduction}
\end{center}

In the  past, a number of observations on the CMB power spectra
\cite{map1,map3,other} and on the large scale structure (LSS)
\cite{lss} have supported inflation as the  good
phenomenological model
in the sense that it naturally gives rise to the origin of the
primordial fluctuations with a nearly scale-invariant and gaussian
spectrum.
The overall expansion of the Universe at very early stage,
as well as the evolution of fluctuations of the perturbed spacetime metric,
can be accounted for in the framework of inflationary models.
In addition to the primordial density perturbations, inflationary
models also predict a stochastic background of RGW,
which is the tensorial  perturbations of spacetime metric.
The detection of such a background  would
provide incontrovertible evidence that inflation actually occurred
and would also set strong constraints on the dynamic of inflation
\cite{sasaki}.

There are mainly two kinds of experiments to detect  RGW at
correspondingly different frequencies. For RGW of very low frequencies
$\nu\sim 10^{-17}\sim10^{-15}$Hz, one can observe them by detecting the
power spectrum of CMB B-polarizations  \cite{B-Pol}.
Now, the first-three-year results of WMAP \cite{map3} have not yet found
the evidence of RGW.
The experiment of the Planck satellite
\cite{planck} with higher sensitivity to polarizations is scheduled
for launch in $2007$, and the Clover (Cl-Observer) \cite{clover}
and CMBPol \cite{CMBPol} projects with much higher sensitivities
than Planck are also under development. For RGW of  high
frequencies $\nu\simeq10^{-4}\sim10^{4}$Hz, another kind of
experiments apply, i.e. the laser interferometers detectors,
including the current TAMA \cite{tama}, VIRGO \cite{virgo}, LIGO
\cite{LIGO,LIGOII}, and the future LISA \cite{LISA}, ASTROD
\cite{ASTROD},
BBO  \cite{BBO},
and DECIGO
\cite{DECIGO}.
Besides these two kinds,
other methods have also been used to constrain the strength of RGW.
For example the timing studies on the
millisecond pulsars,
which can constrain the amplitude of the
gravitational waves by studying the signal residuals of the
millisecond pulses \cite{timing}.
This method is sensitive to the
waves with frequencies at $10^{-9}\sim10^{-7}$Hz.
The electromagnetic detectors are
based on the processes of resonant responses
of electro-magnetic field to incident gravitational waves
\cite{lifangyu, cruise, Bernard}.
This kind of detectors are aiming at detecting RGW
at very high frequencies $\nu\sim 10^{8}-10^{10}$Hz,
which can only be produced by inflation \cite{damp3}.
The observational results of BBN also can constrain the strength of RGW
\cite{BBN,review,book} at all frequencies.
Although the RGW have not been found yet,
some constraints on it have already been obtained by
these experiments or observations.

This paper is to study the various modifications on
the power spectrum of RGW and
examine the constraints on it from the experiments.
On the spectrum, we will consider the modifications due to
such important effects as
the redshift-suppression by various periods of cosmic expansions,
the current accelerating
expansion, and the damping by free-streaming of neutrinos.
Two methods will  be  applied in this study:
the analytic method and the numerical
method by solving the inflationary flow equations.
After considering all these damping effects,
we will arrive at an analytic formula of the strength of RGW,
which, as  a function of the wavenumber $k$,
depends on the scalar spectral index $n_s$ and the
tensor-scalar ratio $r$.
By taking into account of  the current observational constraints
on $n_s$ and $r$, we will obtain an upper limit
of the strength of RGW $\Omega_g<3.89\times10^{-16}$
at $\nu=0.1$Hz.
From the plots in the  $r-\Omega_g$ plane,
we find that the BBO
experiments can detect the RGW if $r>8.3\times10^{-3}$ is satisfied.
This means that BBO is more sensitive than Planck satellite,
but less than Clover and CMBPol.
But the would-be ultimate DECIGO can even detect for
$r>6.8\times10^{-6}$,
much more sensitive than  the CMB experiments.
In this analytic method, $\Omega_g$ depends explicitly on the ratio $r$,
an undetermined parameter, whose
value varies for the various specific inflationary models.
Furthermore,  the approximate power law of
primordial power spectrum also may yield fairly large error
if it is extended over a larger range of frequencies.

To overcome these shortcomings,
we move on to the numerical  method.
Because the RGW depends more sensitively on inflationary stage,
during which it is generated,
and there are a number of inflationary models,
to take care of the predictions from these models,
the numerical method is used,
by which a great many of realizations are produced,
representing the respective inflationary models.
The inflationary flow equations are applied to numerically calculate
the RGW,
whereby an upper limit $\Omega_g<8.62\times10^{-14}$ is obtained
for any slow-roll single-scalar-field inflationary model,
independent of any inflationary parameters.
By taking into account of the current observed constraints
on $n_s$, $\alpha$,  and $r$,
 a much tighter limit $\Omega_g<2\times10^{-17}$
 is arrived at $r\simeq0.03$,
which is beyond the sensitive range of BBO.
In numerically generating $10^7$ realizations,
we find all of them satisfy the current
constraints on $\Omega_g$ from LIGO, from pulsar timing,  and from BBN,
but
only nearly $0.05\%$ of them satisfy the current constraints on
$n_s$, $\alpha$ and $r$.
From the resulting $r-\Omega_g$ plane, one finds
the  DECIGO, if put into running,
will be effective for detecting the RGW with
smaller $r$,
but the CMB experiments, such as Planck, Clover and CMBPol,
are more effective for detecting
the RGW with larger $r$.
They are complementary to each other for  RGW detections.
Our result from this numerical investigation
applies only to the single-field inflationary models
with the chosen initial conditions of Hubble
slow-roll parameters.

The organization of this paper is as follows.
Section 2 gives a simply review on the RGW and its evolution equation.
In section 3, an analytic expression of the strength of RGW
will be obtained with
the three damping factors being included,
presenting the modifications due to the mentioned effects.
Section 4 is devoted to the numerical computations,
where the strength of RGW is numerically calculated
by solving the inflationary flow equations.
Finally section 5 is the conclusion.

~

~

\begin{center}
{\large 2. The relic gravitational waves and their evolutive
equation}
\end{center}
Incorporating the perturbations to the spatially flat
Robertson-Walker (FRW) spacetime, the metric is
 \be\label{1}
 ds^2=a(\tau)^2[d\tau^2-(\delta_{ij}+h_{ij})dx^idx^j]~,
 \ee
where $a$ is the scale factor of the universe, $\tau$ is the
conformal time, which is related to the cosmic time by $ad\tau\equiv
dt$. The perturbation of spacetime $h_{ij}$ is a $3\times3$
symmetric matrix. The gravitational wave field is the tensorial
portion of $h_{ij}$, which is transverse-traceless
$\partial_ih^{ij}=0$, $\delta^{ij}h_{ij}=0$. Since RGW is very
weak, $|h_{ij}|\ll1$, one needs just study the linearized
evolutive equation:
 \be\label{h_evo_0}
 \partial_{\mu}(\sqrt{-g}\partial^{\mu}h_{ij})=16\pi
 Ga^2(\tau)\Pi_{ij}~,
 \ee
where $\Pi_{ij}$ is the tensor part of the anisotropy stress,
satisfying $\Pi_{ii}=0$, and $\partial_i\Pi_{ij}=0$.
It couples to $h_{ij}$ as an external source.
In the cosmic setting, $\Pi_{ij}$ can be generated by
the free-streaming relativistic
particles \cite{damp2,boy}, the cosmic magnetic \cite{magnetic},
etc.
It is convenient to Fourier transform these quantities as follows
 \be\label{3}
 h_{ij}(\tau,{\bf x})=\sum_\lambda\sqrt{16\pi G}
 \int \frac{d~{\bf k}}{(2\pi)^{3/2}}
 \epsilon_{ij}^{(\lambda)}({\bf k})h_{\bf k}^{\lambda}(\tau)e^{i{\bf
 kx}}~,
 \ee
 \be\label{4}
 \Pi_{ij}(\tau,{\bf x})=\sum_\lambda\sqrt{16\pi G}
 \int \frac{d~{\bf k}}{(2\pi)^{3/2}}
 \epsilon_{ij}^{(\lambda)}({\bf k})\Pi_{\bf k}^{\lambda}(\tau)e^{i{\bf
 kx}}~,
 \ee
where the index $\lambda=``+"$ or $``\times"$ labels the two polarization
states of the gravitational waves.
The polarization tensors $\epsilon_{ij}^{(\lambda)}$ are
symmetry, transverse-traceless
$k^i\epsilon_{ij}^{(\lambda)}({\bf
k})=0$, $\delta^{ij}\epsilon_{ij}^{(\lambda)}({\bf k})=0$, and
satisfy the conditions $\epsilon^{(\lambda)ij}({\bf
k})\epsilon_{ij}^{(\lambda')}({\bf k})=2\delta_{\lambda\lambda'}$
and $\epsilon_{ij}^{(\lambda)}({\bf
-k})=\epsilon_{ij}^{(\lambda)}({\bf k})$.
Since the RGW is assumed to be isotropy
 and each polarization state has the same evolution and
 gives the same contributions,
 $h_{\bf k}^{(\lambda)}(\tau)$ is denoted simply by $h_k(\tau)$,
 and $\Pi_{\bf
k}^{(\lambda)}(\tau)$ by $\Pi_k(\tau)$,
where $k=|\textbf{k}|$ is the wavenumber related to the
frequency by $\nu\equiv k/2\pi$ (the present scale factor is set
$a_0=1$).
Then  Eq.(\ref{h_evo_0}) can be rewritten as
 \be\label{h-evolution}
 \ddot{h}_{k}+2\frac{\dot{a}}{a}\dot{h}_k+k^2h_k=16\pi
 Ga^2(\tau)\Pi_k(\tau)~,
 \ee
where the overdot denotes a conformal time derivative $d/d\tau$.
Usually the interactions  between gravitational waves and other
matters are very weak, in many cases,
the source $\Pi_k$ in Eq.(\ref{h-evolution}) is negligible,
and the evolution of RGW only depends on the scale factor
and its time derivative.
But in this paper we include this source term,
so that the damping from neutrino free-streaming is properly
taken care of.

~

~

\begin{center}
{\large 3. The analytic power spectrum of RGW }
\end{center}
\emph{\textbf{The primordial power spectrum of RGW}}\\
As mentioned in the introduction,
inflationary expansion, as an attractive idea to describe the very
early universe,  has received strong support from the
observations of CMB anisotropies and from studies of the
large-scale distribution of galaxy.
There have been a number of models proposed.
Here we will consider only the single field models.
In the context of
slow-roll inflationary models, the  observables depend on
three slow-roll parameters \cite{slow-roll}
 \be\label{p}
 \epsilon_V\equiv\frac{\Mpl^2}{2}\left(\frac{V'}{V}\right)^2~,
 ~~~~~~
 \eta_V\equiv \Mpl^2\left(\frac{V''}{V}\right)~,
 ~~~~~~
 \xi_V\equiv \Mpl^4\left(\frac{V'V'''}{V^2}\right)~,
 \ee
where $\Mpl\equiv(8\pi G)^{-1/2}=\mpl/\sqrt{8\pi}$ is the reduced
Planck energy,
 $V(\phi)$ is the inflationary potential, and the prime
denotes derivatives with respect to the field $\phi$.
Here,
$\epsilon_V$ quantifies ``steepness" of the slope of the
potential, $\eta_V$ measures ``curvature" of the potential,
and $\xi_V$ quantifies the ``jerk".
All these three parameters must be smaller
than $1$  for inflation to occur.
One of the important predictions of
 inflationary models is the primordial scalar perturbation
power spectrum, which is nearly gaussian and nearly
scale-invariant.
This spectrum can be  written in the form
 \be\label{s-p}
 P_S(k)=P_S(k_0)\left(\frac{k}{k_0}\right)^{n_s(k_0)-1
 +\frac{1}{2}\alpha\ln(k/k_0)}~,
 \ee
where $n_s$ is the scalar spectral index,
 $\alpha\equiv dn_s/d\ln k$ is
its running, and $k_0$ is some pivot wavenumber.
In this paper,  $k_0=0.05$Mpc$^{-1}$ is taken,
which corresponds to a frequency $\sim 10^{-16}$Hz.
The observations of WMAP gives $P_S(k_0)\simeq2.95\times10^{-9}A(k_0)$ and
$A(k_0)=0.9\pm0.1$ \cite{map1}.
Another major prediction of
inflationary models is the existence of  RGW.
The primordial power spectrum of RGW is defined by
 \be\label{6}
 P_T(k)\equiv\frac{32Gk^3}{\pi}h^{+}_kh_k~,
 \ee
where $h_k$ is the solution of the equation in (\ref{h-evolution}).
This spectrum can also be put in a simple
form
 \be\label{7}
 P_T(k)=P_T(k_0)\left(\frac{k}{k_0}\right)^{n_t(k_0)+\frac{1}{2}\alpha_t\ln(k/k_0)}~,
 \ee
where $n_t(k)$ is the tensor spectral index, and $\alpha_t\equiv
dn_t/d\ln k$ is its running.
In the single-field inflationary models,
a standard slow-roll analysis gives the following relations
 \be\label{relation}
 n_t=-\frac{r}{8}~,
 ~~~~~~
 \alpha_t=\frac{r}{8}\left[\left(n_s-1\right)+\frac{r}{8}\right]~,
 ~~~~~~
 r=\frac{8}{3}(1-n_s)+\frac{16}{3}\eta_V~,
 \ee
where $r(k)\equiv P_T(k)/P_S(k)$ is the so-called tensor-scalar
ratio.
These formulae relate the tensorial parameters $n_t$ and $\alpha_t$ to
the scalar parameters $n_s$ and $r$;
the latter are accessible  to the observations of CMB and of LSS.
As shown in Eq.(\ref{relation}),
the relation between $r$ and $n_s$ involves the slow-roll parameter $\eta_V$,
depending on the specific inflationary potential.
Inserting these
into Eq.(\ref{7}), one has
 \be\label{pt1}
 P_T(k)=P_S(k_0)\times r\times\left(\frac{k}{k_0}\right)^
 {-\frac{r}{8}+\frac{r}{16}\left[\left(n_s-1\right)+\frac{r}{8}\right]\ln(k/k_0)}~.
 \ee
In general,  the tensor-scalar ratio $r$ may vary with the wavenumber $k$.
Here and in the following sections
we will take the value of $r$  at  $k=k_0$, i.e.
$r\equiv r(k_0)$.
Now the primordial spectrum of RGW only depends
on the parameters $n_s$ and $r$.
The recent constraints by the observations of three-year
WMAP, SDSS, SNIa and galaxy clustering \cite{se} are
 \be\label{sdss}
 n_s=0.965\pm0.012~,~~~~(68\% ~C.L.)
 \ee
 \be
r<0.22~,  ~~~(95\% ~C.L.)~.
 \ee
From the derivation it is obvious that
the formula (\ref{pt1}) applies properly for in a frequency range
around $\sim 10^{-16}$Hz.
The strength of the gravitational waves can be also characterized by the
gravitational waves energy spectrum
 \be\label{26}
 \Omega_{g}(k)=\frac{1}{\rho_c}\frac{d\rho_{g}}{d\ln k}~,
 \ee
where $\rho_c=3H_0^2/8\pi G$ is the critical density and
$H_0=100h~$km s$^{-1}$Mpc$^{-1}$ is the present Hubble constant
(the value $h=0.72$ is taken through this paper).
$\Omega_g$ can be related to the primordial power spectrum by the formula
\cite{boy,before-work}
 \be\label{ome}
 \Omega_{g}(k)=\frac{1}{12H_0^2}k^2P_T(k)T^2(k)~,
 \ee
where the transfer function $T(k)$ will take into account of
 the  various damping effects mentioned early, and
 will be discussed below.

~\\
\emph{\textbf{The damping effects}}\\
Here  three kinds of damping effects will be addressed:
First we only consider the redshift-suppression effect
caused by the overall expansions of the universe.
So temporarily we drop the the anisotropy stress
term  $\Pi_k(\tau)$  in Eq.(\ref{h-evolution})
due to neutrino free-streaming,
 \be\label{vacuum}
 \ddot{h}_{k}+2\frac{\dot{a}}{a}\dot{h}_k+k^2h_k=0~.
 \ee
This equation of RGW only
depends on the behavior of the scale factor $a(\tau)$.
 It has been known that, during the expansion of the universe,
the mode function $h_k(\tau)$ of the gravitational waves behaves
differently in two regimes \cite{damp3}: far outside
the horizon ($k\ll aH$), and deep inside the horizon ($k\gg aH$).
When waves are far outside the horizon, the amplitude of $h_k$
keeps constant, and when inside the horizon, the amplitude is damping
with the expansion of the universe
 \be\label{28}
 h_k\propto \frac{1}{a(\tau)}  ~~.
 \ee
By numerically integrating the Eq.(\ref{vacuum}),
this effect can be
approximately described by  a transfer function
\cite{damp1}
 \be\label{29}
 t_1(k)=\frac{3j_1(k\tau_0)}{k\tau_0}
 \sqrt{1.0+1.36\left(\frac{k}{k_{eq}}\right)+2.50\left(\frac{k}{k_{eq}}\right)^2}~~,
 \ee
where $k_{eq}=0.073\Omega_mh^2$Mpc$^{-1}$ is the wavenumber
corresponding to the Hubble radius at the time that matter and
radiation have equal energy densities. And
$\tau_0=1.41\times10^{4}$Mpc is the present conformal time.
It is obvious that,
this factor $t_1(k)$ is oscillating with wavenumber $k$
due to the Bessel function $j_1(k\tau_0)$.
In practice, one is usually interested in the overall outline of
the amplitude of RGW,
as the quick oscillations with $k$ are of no importance.
For the waves with $k\tau_0\gg1$,
this factor can be written as
\be\label{t1}
t_1(k)=\frac{3}{(k\tau_0)^2}
 \sqrt{1.0+1.36\left(\frac{k}{k_{eq}}\right)
 +2.50\left(\frac{k}{k_{eq}}\right)^2}~~.
\ee

The above transfer function (\ref{t1}) does not include the
effect of accelerating expansion of the present universe, which has
been indicated by the  observations on SN 1a.
The spectrum of
RGW has been studied in specific models for dark energy
\cite{accele}, such as the Chaplyngin gas models and the X-fluid model.
In the Ref.\cite{damp3}, we have presented an analytic solution
of RGW in $\Lambda$CDM universe,
and found that the  amplitude
of the gravitational waves has been modified by the presence of the
dark energy during  the current expansion.
In the higher frequency range
($\nu\gg3\times10^{-18}$Hz) that we are interested in this paper, the
amplitude acquires an overall factor $\Omega_m/\Omega_{\Lambda}$
as compared with the decelerating model, where $\Omega_m$ and
$\Omega_{\Lambda}$ are the present energy densities of matter and
vacuum, respectively.
So this effect can be simply described by a
damping factor,
 \be\label{t2}
 t_2=\frac{\Omega_m}{\Omega_{\Lambda}}~.
 \ee
In the standard $\Lambda$CDM model with $\Omega_m=0.27$ and
$\Omega_{\Lambda}=0.73$,
this effect contributes a damping factor
of $t_2^2\sim 0.137$ for the strength of RGW in Eq.(\ref{ome}).

The third to be  considered  is the damping effect of the
free-streaming neutrinos  \cite{damp2}, i.e. the anisotropic
stress $\Pi_k$ on the right-hand of the Eq.(\ref{h-evolution}).
This effect has been considered by Weinberg \cite{damp2}.
This effect is primarily produced by  neutrinos
during the radiation dominated era when they are
decoupled and are  free streaming in the universe, especially
right after the gravitational waves enter the the horizon.
The overall amplitude
of RGW will reduce roughly by an amount of $20\%$.
It has been shown that
anisotropy stress can reduce the amplitude for the wavelengths
that re-enter the horizon during the radiation-dominated stage,
and the damping factor is only dependent on the fraction $f$ of
the free-streaming relativistic particles over the background
(critical) energy density in the universe.
For the waves that enter the horizon at later times,
the damping effect is less important.
A number of
works have done to discuss this effect, and  Reference \cite{boy}
has found that the effect can be approximately described by a
transfer function $t_3$ for the waves with wavenumbers
$k>10^{-16}$Hz (which re-enter the horizon at the
radiation-dominant stage),
 \be\label{t3}
 t_3=\frac{15(14406f^4-55770f^3+3152975f^2-48118000f+324135000)}
 {343(15+4f)(50+4f)(105+4f)(180+4f)}~~.
 \ee
When the wave modes ($10^{-16}$Hz$<k<10^{-10}$Hz) re-enter the
horizon, the temperature in the universe is relatively low
($<1$MeV), the neutrino is the only free-streaming relativistic
particle.
Thus we choose the fraction $f=0.4052$,
corresponding to $3$ species of neutrinos in the standard model,
the damping factor is then $t_3= 0.80313$. But for
the waves with very high frequency ($k>10^{-10}$Hz), the
temperature of the universe is still very high when they re-enter the
horizon, the neutrinos were still in interaction,
and the value of $f$ is quite uncertain. This is because
the detail of how many species of particles are free is not
accurately known and depends on the cosmic environment
and the particle-interaction models.
Thus, the detection of RGW at this range of frequencies
offers the possibility of learning more about the free-streaming
fraction $f$ in the very early universe.
So we choose $f=0$, i.e. there are no
free-streaming relativistic particle at that time.
The corresponding factor is then $t_3=1$.
But for the waves with $k<10^{-16}$Hz, which re-enter the
horizon during the matter-dominated stage,
the neutrino density is so
small that its damping impact can be neglected, so we choose
$t_3=1$.
Overall, the neutrino damping effect can approximately
be summarized as
 \be t_3(k)\simeq \left\{
 \begin{array}{ll}
 1, & k<10^{-16}Hz \\
 0.80313, & 10^{-16}Hz<k<10^{-10}Hz\\
 1, & k>10^{-10}Hz.
 \end {array}
 \right. ~~~
 \ee

Putting these three effects together,
the total transfer function is
the combination of three factors
 \be\label{t}
 T(k)=t_1\times t_2 \times t_3~,
 \ee
among these
$t_1$ is dominantly  important, which approximately shows the
evolution of RGW in the expanding universe.
The function $t_2$ has
the relatively smaller damping on RGW,
which accounts for the accelerating expansion
of the universe quite recently ($z\sim 0.3$).
The value of energy spectrum
$\Omega_g$ is reduced by nearly an order by this effect.
The function of $t_3$ has the most uncertainty,
 as has been discussed above.
In the extreme case with $f=0$, one has $t_3=1$, i.e. no
damping; and in another extreme case with $f=1$,  one has $t_3=0.35$,
the smallest value of $t_3$.
In the case of $f=0.4052$, $t_3=0.80313$,
 only contributing a damping factor
$t_3^2 = 0.645$ for the strength of the RGW.

There are some other possible mechanisms,
which might affect the amplitude of
the gravitational waves.
For example,  the QCD transition
\cite{qcd,trans}, $e^+e^-$ annihilation \cite{qcd,ee,trans}, the
cosmic reheating \cite{grishchuk,damp3},  and so on \cite{boy}.
These could influence the value of expansion rate $\dot{a}/a$,
 and therefore affect the strength of RGW.
However,  these effects are either small, as shown in literature,
 in comparison with the effects we have discussed,
or there are some  uncertainties in their analysis,
so these are not considered in here.

~\\
\emph{\textbf{ The upper limit of $\Omega_g$ and the sensitivities
of future experiments }}

The future detectors of RGW are mainly classified into two kinds:
one kind is through CMB for very low frequencies,
and another is based on laser interferometers for relatively high frequencies.

For the waves of very low frequencies $\nu<10^{-15}$Hz,
the CMB experiments are sensitive.
For instance,
the Planck satellite can detect the RGW if
$r>0.1$ \cite{planck}, the ground-based experiment Clover can
detect the signal if $r>0.005$ \cite{clover}, and CMBPol can
detect if $~r>10^{-3}$ is satisfied \cite{CMBPol}. It should
notice that if~ $r<1\times10^{-4}$, the RGW may not be detected by
the CMB experiments.
This is because the CMB B-polarizations generated by the cosmic lensing
are also very large,
and the signals from the RGW may be subdominant to
the lensing effects.
\cite{lensing}.

The  direct detections the RGW by  laser
interferometers  are sensitive to the waves with  high
frequencies.
For the waves with $k>10^{-10}$Hz, inserting the formulas
(\ref{t1})-(\ref{t}) with $t_3=1$ in Eq.(\ref{ome}), the strength of the
gravitational waves becomes
 \be
 \Omega_{g}(k)=\frac{22.5}{12H_0^2}\frac{P_T(k)}{\tau_0^4k_{eq}^2}
 \left(\frac{\Omega_m}{\Omega_{\Lambda}}\right)^2\simeq1.08\times10^{-6}P_T(k)~~,
 \ee
 Using the
expression of $P_T(k)$ in Eq.(\ref{pt1}), one gets
 \be\label{ome1}
 \Omega_{g}(k)\simeq2.87\times10^{-15}~r~\left(\frac{k}{k_0}\right)^
 {-\frac{r}{8}+\frac{r}{16}\left[\left(n_s-1\right)+\frac{r}{8}\right]\ln(k/k_0)}~,
 \ee
where  $A(k_0)=0.9$ has been taken.
This function depends on the
wavenumber $k$, the tensor-scalar ratio $r$ and the scalar
spectral index $n_s$.
It should be pointed out that,
just as Eq.(\ref{pt1}),
this formula (\ref{ome1}) also
applies properly in a frequency range around  $\sim 10^{-16}$Hz.

The advanced LIGO can detect the waves with~
$\Omega_{g}h^2>10^{-9}$ at $\nu\simeq100$Hz \cite{LIGOII}.
The LISA project is expected to detect waves with $\Omega_{g}h^2>10^{-11}$ at
$\nu\simeq0.005$Hz \cite{LISA}.
The ASTROD, a space project sensitive to the waves with frequency at
$\nu\in( 10^{-5},~10^{-3})$Hz \cite{ASTROD},
is expected to detect the waves with $\Omega_{g}h^2>10^{-15}$ at
$\nu\simeq5\times 10^{-4}$Hz.
The BBO, another important project,
 can detect a background RGW with
$\Omega_{g}>2.2\times 10^{-17}$ at $\nu\simeq(0.1$-$1)$Hz
\cite{BBO}. The DECIGO project,  having a much higher
sensitivity by design, is expected to detect the RGW
with $\Omega_{g}h^2>10^{-20}$ at $\nu\simeq0.1$Hz \cite{DECIGO}.

First, we will estimate the upper limit on the strength of RGW in
Eq.(\ref{ome1}).
Here we assume $n_s\leq1$ and $r<0.22$,
which are consistent with the current observations
\cite{se}.
The formula (\ref{ome1}) gives an upper limit of
$\Omega_g$ at $\nu=0.1$Hz:
 \be\label{limit0}
 \Omega_g<3.89\times10^{-16}~.
 \ee
And this limit is arrived at $n_s=1$ and $r=0.22$. This limit is
nearly an order smaller than the result in
Ref.\cite{before-work}.
This is because our analysis has taken into account of
the damping effect of the accelerating expansion of the universe and
the running of $n_t$ in the primordial spectrum.
This limit is in
the sensitive ranges of BBO and DECIGO,
but beyond those  of LIGO, LISA and ASTROD.
In Fig.[1], we plot the strength of RGW at
$\nu=0.1$Hz, as the function of $r$, where  several
models with different $n_s$ are demonstrated.
One sees that when $r<0.01$, the curves of the function $\Omega_g$
are almost overlapped  for
the models of different $n_s$,  and only depend on the variable $r$.
But when $r>0.01$, the models of different $n_s$ can be distinguished.
 For a fixed $r$, a larger $n_s$ yields a larger $\Omega_g$.
This figure also tells that
BBO can detect the RGW if $r>8.3\times 10^{-3}$,
so it is more sensitive than the Planck satellite, but
less than Clover and CMBPol.
It is interesting to notice that
DECIGO can detect the RGW if $r>6.8\times 10^{-6}$, which is much
more sensitive than all the CMB experiments ($r>10^{-4}$).

~\\
\emph{\textbf{The predictions of inflationary models}}

The strength of RGW in Eq.(\ref{ome1}) depends on the values of
$n_s$ and $r$.
Observations have yielded quite solid constraints on
$n_s$, but the value of $r$ is still uncertain.
The
relation between $n_s$ and $r$ depends on the specific
inflationary models,
and different  models will predict very different
$r$.
In the following  examinations will made for
several inflationary models,
which predict different values of $r$.
One may
categorize slow-roll models into several classes according to
the parameter space spanned by $n_s$,
$\alpha$ and $r$ \cite{cata}.
Each class should correspond to
specific physical models of inflation.
Here we categorize the
models according to the curvature of potential $\eta_V$ in Eq.(\ref{p}),
as it is the
only parameter that enters into the relation (\ref{relation})
between $n_s$ and $r$.
 The classes are defined in the following:

\emph{Case A: negative curvature models $\eta_V<0$}

The negative $\eta_V$ models often arise from a potential of
spontaneous symmetry breaking.
One type of often-discussed  potentials have
the form of $V=\Lambda^4\left[1-(\phi/\mu)^p\right]$, where
$p\geq2$. This kind of models predict the red tilt $n_s<1$, which
is consistent with the observations of three-year WMAP.
Also these
models predict pretty  small $r$.
For the model with $p=2$ in Ref.\cite{cata},
 \be\label{casea}
 r\simeq 8(1-n_s)e^{-N(1-n_s)}~,
 \ee
where $N$ is the number of e-folds, taken be in the
range $N\in[40,~70]$ to account for the current observations on
CMB~\cite{N,map1,map3}. Here we choose the value $N=70$.
Using the constraint on $n_s$ in Eq.(\ref{sdss}) yields
the constraint $r\in[0.014,~0.037]$. From Fig.[1], one finds this
is beyond the sensitive range of the Planck satellite, but in the
sensitive ranges of Clover, and CMBPol. And it is also in the
sensitive range of BBO and  DICIGO. In other models with
$p>2$, the predicted values of $r$ are much smaller than
that of the model with $p=2$.

\emph{Case B: small positive curvature models
$0\leq\eta_V\leq2\epsilon_V$}

These models contain as two examples the monomial potentials
$V=\Lambda^4(\phi/\mu)^p$ with $p\geq2$ for
$0<\eta_V<2\epsilon_V$ and the exponential potential
$V=\Lambda^4\exp(\phi/\mu)$ for $\eta_V=2\epsilon_V$. In these
models, to the first order in slow roll, the scalar index is
always red $n_s<1$ and the following constraint on $r$ is
satisfied
 \be\label{caseb}
 \frac{8}{3}(1-n_s)\leq r\leq8(1-n_s)~.
 \ee
Using the constraint on $n_s$ in Eq.(\ref{sdss}), one finds that
$r\in[0.061,~0.376]$, which is in the sensitive ranges of Clover,
CMBPol, BBO,  and DECIGO.
The sensitivity limit of Planck is just in this span,
so it may be able to detect the model.

\emph{Case C: intermediate positive curvature models
$2\epsilon_V<\eta_V\leq3\epsilon_V$}

The supergravity-motivated
hybrid models have a potential of the  form
$V\simeq\Lambda^4\left[1+\alpha\ln(\phi/Q)+\lambda(\phi/\mu)^4\right]$,
up to one-loop correction
 during inflation.
In this case,
 \be\label{cased}
 n_s<1~,~~~~~~r>8(1-n_s)~,
 \ee
are satisfied. Using the constraint on $n_s$ in Eq.(\ref{sdss}),
one finds that $r>0.184$, which is very close to the current upper
limit $r<0.22$.
Fig.[1] shows that this model is in the sensitive
range of Planck satellite.

\emph{Case D: large positive curvature models
$\eta_V>3\epsilon_V$}

This class of models have a typical monomial potential similar to the Case A,
but with  a  plus sign for the term $(\phi/\mu)^p$:
$V=\Lambda^4\left[1+(\phi/\mu)^p\right]$.
This will enable inflation to occur for a small value of $\phi<\mpl$.
 This model
predicts a blue tilt of scalar index $n_s>1$, which is contradict
to the constraint in Eq.(\ref{sdss}).
But we should notice that
the observations of three-year WMAP has not yet ruled out the blue
spectrum.
If  the the running of $n_s$ with the wavenumber $k$ is allowed,
 then the best fit of
WMAP data suggests that
$n_s(k=0.002$Mpc$^{-1})=1.21^{+0.13}_{-0.16}$, and
$\alpha(k=0.002$Mpc$^{-1})=-0.102^{+0.050}_{-0.043}$\cite{map3}.
This is a blue spectrum with a negative running. So the
determination of the value of $n_s$ depends on the more precise
observations.

~

~

\begin{center}
{\large 4. The inflationary flow equations and the predictions for
RGW}
\end{center}
In the discussions, the RGW given by the analytic expression
(\ref{ome1}) depends on the value of tensor-scalar ratio $r$,
which  has not yet been determined by the observations. Moreover,
Eq.(\ref{ome1}) is a  good approximation only for the waves with
wavenumber around $k\simeq k_0(\sim 10^{-16}$Hz).
Therefore, for the waves of high frequencies,
say  with $\nu  \simeq 0.1$Hz,
nearly $15$ orders larger than
the value of $k_0 $,
direction application of the formula (\ref{ome1})
may lead to  large errors.
Consequently, it will be restricted in practice.
To
avoid these shortcomings, in this section, we will employ the technique
of the inflationary flow equations to relate RGW in lower
frequencies to that in higher frequencies.

~\\
\emph{\textbf{The inflationary flow equations}}

The inflationary flow equations were first introduced by Hoffman
and Turner \cite{flow} as a way of generating a large number of
slow-roll inflationary models to be compared to the observational
data. This method applies to any slow-roll single scalar field
inflationary models, and relies on defining a set of Hubble
slow-roll parameters, which are the derivatives of the Hubble
parameter during inflation. The major advantage of this method is
that it removes the field from the dynamics, and allows one to
study the generic behavior of slow-roll inflation without making
detailed assumptions about the underlying particle physics. In
this section, we will also use this method to generate a large
number of inflationary models, the observables of which are required
to be  consistent
with the current observational constraints in low frequencies.
Then we will numerically solve the strength of RGW in  very high frequencies.
In this method the Hubble slow-roll parameters are defined by
 \be\label{8}
 \epsilon(\phi)\equiv\frac{\mpl^2}{4\pi}\left(\frac{H'(\phi)}{H(\phi)}\right)^2~,
 ~~~~~~
 \lambda_l(\phi)\equiv\left(\frac{\mpl^2}{4\pi}\right)^l\frac{(H')^{l-1}}
 {H^l}\frac{d^{(l+1)}H}{d\phi^{(l+1)}}, ~ ~~~
 (l\geq1)~,
 \ee
where primes are derivatives with respect to the scalar field  $\phi$, and
$H(\phi)$ is the Hubble parameter as the function of $\phi$,
related to the potential $V(\phi)$ by the
so-called Hamilton-Jacobi formula,
 \be\label{hj}
 \left[H'(\phi)\right]^2-\frac{12\pi}{\mpl^{2}}H^2(\phi)
 =-\frac{32\pi^2}{\mpl^{4}}V(\phi)~.
 \ee
These Hubble slow-roll parameters satisfy an infinite set of
hierarchical differential equations, called inflationary flow equations:
 \be\label{10}
 \frac{d\epsilon}{d N}=\epsilon(\sigma+2\epsilon)~,
 \ee
 \be\label{11}
 \frac{d\sigma}{d N}=-\epsilon(5\sigma+12\epsilon)+2( \lambda_2)~,
 \ee
 \be\label{12}
 \frac{d}{d N} \lambda_l=\left[\frac{l-1}{2}\sigma+(l-2)\epsilon\right]
 \lambda_l+ \lambda_{l+1}~,~(l\geq2)
 \ee
where $N$ is the number of $e$-folds of the inflation, and
$\sigma\equiv2 \lambda_1 -4\epsilon$. There are two families of
fixed points of these flow equations: One is that $\epsilon=0$,
$\lambda_l=0$ for $l\ge 2$,  and $\sigma$=constant. In
Ref.\cite{attractor}, the authors found that, only if $\sigma>0$,
this fixed point is stable, i.e. the attractor solution. The other
family of fixed points are given by: $\epsilon$=constant,
$\sigma=-2\epsilon$, $\lambda_2=\epsilon^2$, and
$\lambda_l=\epsilon\lambda_{l-1}$ for $l\ge 3$.
Later we will
show  that the second family of  fixed points is not stable. The
slow-roll parameters tend to run to the attractor with the
expansion of the universe, as long as the slow-roll condition
$\epsilon<1$ is satisfied. In order to actually solve this
infinite series of equations numerically,
it must be truncated at some $l$ by
setting a sufficiently high slow-roll parameter to zero, i.e.
$\lambda_{m+1}$=b, with $b$ being a constant,
 and $\lambda_{m+2}$=$0$ for some
suitably large $m$.
In this section, we make the truncation of   this series at
$m$=$10$,
and choose a set of acceptable initial conditions as in
Refs.\cite{attractor,before-work}:
 \be\label{13}
 \epsilon|_i\in[0,0.8]~,
 \ee
 \be\label{14}
 \sigma|_i\in[-0.5,0.5]~,
 \ee
 \be\label{15}
 \lambda_{2}|_i\in[-0.05,0.05]~,
 \ee
 \be\label{16}
 ~~~~~~~~~~~~~~~~~\lambda_{l}|_i\in[-0.025\times 5^{-l+3},0.025\times 5^{-l+3}]~,
              ~~~(3\leq l\leq10)~,
 \ee
 where the subscript  $|_i$ denote the corresponding initial values.
This set of eleven equations in
Eqs.(\ref{10})-(\ref{12}) is an autonomous system
\cite{attractor}.
We choose the constant  $b\neq$0, and set the left hand of these
equations to be zero.
Then  the only real solution
for this eleven-equation set is given by
 \be\label{17}
 \epsilon_c=b^{1/11}~,~ \sigma_c=-2b^{1/11}~,~
 \lambda_{lc}=b^{l/11}, ~~ (2\leq
 l\leq10)~,
 \ee
 where the subscript $c$ means the fixed point.
This is just the second family of fixed points with
$\epsilon_c=b^{1/11}$.
As usual,
in order to study the stability of this
fixed points, let us consider the small perturbations
around the fixed point, i.e.
 \be
 \epsilon=\epsilon_c+\delta\epsilon,~\sigma=\sigma_c+\delta\sigma,
 ~\lambda_{l}=\lambda_{lc}+\delta\lambda_{l}, ~~~(2\leq
 l\leq10)~.
 \ee
Substituting these into Eqs.(\ref{10})-(\ref{12}), one gets the
first-order differential equations for the small perturbations
 \be
 \frac{d}{d N}\left(
 \begin{array}{ccc}
    \delta\epsilon\\
    \delta\sigma\\
     .\\
     .\\
    \delta\lambda_{10}
 \end{array}
 \right)=M\left(
 \begin{array}{ccc}
    \delta\epsilon\\
    \delta\sigma\\
     .\\
     .\\
    \delta\lambda_{10}
 \end{array}
 \right)~~,
 \ee
where the matrix $M$ depends upon the values of $\epsilon_c$,
$\sigma_c$ and $\lambda_{lc}$, $l=2,..., 10$.
If this fixed point is stable, at
least, it is necessary that the real parts of the eigenvalues of
the matrix  $M$ are negative \cite{de}.
We have performed calculations and found that
no matter what value of $b\ne 0$ is chosen,
this condition can not be
satisfied.
Therefore, this fixed point is not stable.
If we choose the value $b=0$, this eleven-equation set in
Eqs.(\ref{10})-(\ref{12}) has fixed points,
which belong to the first family mentioned early
and are stable only if $\sigma_c >0$ is satisfied.

It is obvious that the
evolutions of this eleven-equation set will be  different for the
conditions with different $b$.
Although whether the fixed points are stable or not depends on
the value of $b=0$,
still in the computation below, we
will do computations for both kinds of initial conditions: one
with $b=0$, and the other $b\ne 0$ with
 \be
 b\in[-0.025\times 5^{-8},0.025\times 5^{-8}]~.
 \ee
It will turn out that, for these two cases,
our calculational results of $\Omega_g$
are very similar.
So in the following sections,
we will only present  the results of the   $b=0$ case.

~\\
\emph{\textbf{The inflationary parameters and the strength of
RGW}}

Many observable parameters in  inflationary models can be
related to the . Here we are only
interested in three such kind of observable parameters for the slow-roll
inflationary models: the tensor-scalar ratio $r$, the scalar
spectral index $n_s$, and its running $\alpha$.
They are related to the Hubble slow-roll parameters
as the following  (up to the second order in the slow-roll) \cite{para}
 \be\label{19}
 r\simeq16\epsilon[1-c(\sigma+2\epsilon)]~,
 \ee
 \be\label{20}
 n_s\simeq1+\sigma-(5-3c)\epsilon^2
     -\frac{1}{4}(3-5c)\sigma\epsilon+\frac{1}{2}(3-c)\lambda_2~,
 \ee
 \be\label{22}
 \alpha=-\frac{1}{1-\epsilon}\frac{dn_s}{d N}~,
 \ee
where $c=4(\ln 2+\gamma)-5\simeq0.0814514$ (with $\gamma$ the
Euler-Mascheroni constant) is a constant.
Once the inflationary flow equations in
(\ref{10})-(\ref{12}) are
numerically solved, the values of these three observables  are obtained.
As we have pointed out early,
 we are interested
for the gravitational waves in a very wide  frequency range,
$\nu\in[10^{-16},~10^2]$Hz,
and the
primordial power spectrum in Eq.(\ref{7}),
as an analytic approximation,  may not apply properly.
Therefore, we need to adopt
the following primordial power spectrum\cite{lyth}
 \be\label{23}
 P_T(k)=\left.\frac{16}{\pi}
 \left[1-\frac{c+1}{4}\epsilon\right]^2\frac{H^2}{\mpl^2}\right|_{k=aH}~,
 \ee
which is proper for the slow-roll inflationary models in general.
Here $H$ is the Hubble parameter of inflation when the waves
exactly crossed the horizon with $k=aH$.
If one ignored the small slow-roll parameter $\epsilon$ in Eq.(\ref{23}),
then  one would end up with
$P_T(k)=\left.\frac{16}{\pi} \frac{H^2}{\mpl^2}\right|_{k=aH}$,
a result for the exact de Sitter inflation,
depending only on the Hubble parameter $H$.
The formula (\ref{23}) can be rewritten as
 \be\label{24}
 P_T(k)=\left(\frac{4-(c+1)\epsilon}{4-(c+1)\epsilon_i}\right)^2
 \frac{H^2}{H_i^2}P_T(k_0)~,
 \ee
where  $\epsilon_i$  and $H_i$ are the respective values
of $\epsilon$ and $H$  when $k_0$ just crosses the
horizon at $a=k_0/H_i$.
By the way, the RGW power spectrum can be related  to the
scalar one by $P_T(k_0)=P_S(k_0)r(k_0)$ as before.
The value of $H$ is also  related to $H_i$ through
the parameter $\epsilon$ by the following
 \be\label{25}
 H(N)=H_i\exp\left[-\int_{N_i}^N\epsilon(n)d n\right]~,
 \ee
where $N_i$ is the number of e-folds if $H=H_i$.
Inserting the
Eqs.(\ref{24}) and (\ref{25}) into Eq.(\ref{ome}),
one obtains the energy spectrum of RGW
 \be\label{ome2}
 \Omega_g(k)=2.21\times 10^{-10}~r \left(\frac{k }{H_0}\right)^2 T(k)^2
 \left(\frac{4-(c+1)\epsilon}{4-(c+1)\epsilon_i}\right)^2
 \exp\left[-2\int_{N_i}^N\epsilon(n)d n\right]~,
 \ee
where $T(k)$ is the damping factor, and $H_0$ is the present
Hubble constant.
Making use of  $T(k)$ as given in Eqs.(\ref{t1}),
(\ref{t2}), (\ref{t3}) and the ratio $r$ in Eq.(\ref{19}),
one converts the above energy spectrum  of RGW into the following form
 \be \label{main}
 \Omega_{g}(k)
 \simeq4.59\times10^{-14}\left[\epsilon_i-c(\sigma_i\epsilon_i+2\epsilon_i^2)\right]
 \left(\frac{4-(c+1)\epsilon}{4-(c+1)\epsilon_i}\right)^2
 \exp\left[-2\int_{N_i}^N\epsilon(n)d n\right]~,
 \ee
which now depends only on the slow-roll parameters $\epsilon$
and $\sigma$.
As an advantage to Eq.(\ref{ome1}),
this spectrum is good also for the waves with $\nu \gg 10^{-10}$Hz.
Before numerically computing it
through the inflationary flow equations,
we first give an estimate of its upper limit.
Since $0\leq\epsilon<1$
is satisfied during the inflation, Eq.(\ref{main})
yields the upper limit when taking $\epsilon=0$
 \be\label{<}
 \Omega_g(k)<7.34\times10^{-13}
 \frac{\epsilon_i-c(\sigma_i\epsilon_i
 +2\epsilon_i^2)}{\left(4-(c+1)\epsilon_i\right)^2}~,
 \ee
depending on the values of $\epsilon_i$ and $\sigma_i$.
When $\epsilon_i=1$, the right-hand of this
inequality has the maximum value,
so one can give a loose upper
limit of $\Omega_g(k)$:
 \be\label{limit1}
 \Omega_g(k)<8.62\times10^{-14}~,
 \ee
where the approximation $r\simeq16\epsilon$ has been used,
and the second-order terms of the tensor-scalar ratio
$r$  have been omitted.
This upper limit holds only if the slow-roll
condition is satisfied.
Compared with the limit in Eq.(\ref{limit0}),
this upper limit is arrived at
without explicitly  using  the values of $n_s$ and $r$.
Besides, this limit applies for
a wider range of frequencies $\nu >10^{-10}$Hz.
It is seen that this limit is much less stringent
than that in Eq.(\ref{limit0}).
Still the  limit (\ref{limit1})
is much smaller than the sensitivity  of,
and therefore can not be directly detected by  LIGO and LISA,
however, it is now larger than the sensitivity of,
and can be detected by  ASTROD, BBO and DECIGO.
In writing down the limit (\ref{limit1}),
$t_3=1$ has been used,
which is valid for waves with $\nu \gg10^{-10}$Hz.
But for the waves with $\nu \in[10^{-16},~10^{-10}]$Hz,
one should use $t_3\simeq 0.80313$,
and thus the limit becomes
$\Omega_g(k)<5.56\times10^{-14}$, which is a little tighter than
the limit in Eq.(\ref{limit1}).

~\\
\emph{\textbf{The current constraints on the cosmic parameters}}

Through the  discussion above, we know that the values of
the inflationary parameters  $n_s$,
$\alpha$, $r$, and $\Omega_g$ are all directly related to the
Hubble slow-roll parameters.
Here, we give a  review of the
current constraints on them.
The constraints on $n_s$, $\alpha$ and $r$ mainly come from the
observations on large scales, including the observations of CMB,
LSS and so on. Here we call them as ``large-scale constraints"
(LSC).
When solving the inflationary flow equations,
we will take the initial condition at the time of horizon-leaving,
i.e. $k_0=aH$,
where the povit wavenumber $k_0=0.05$Mpc$^{-1}$ as before.

Now  the WMAP CMB data (1st year)
gives \cite{map1} $n_s=0.93\pm0.07$, $\alpha=-0.047\pm0.040$,
and the best fit of WMAPext+2dFGRS galaxy survey gives $n_s=0.93\pm0.03$,
$\alpha=-0.031^{+0.016}_{-0.017}$. A fit using WMAP CMB data and
the SDSS galaxy survey gives \cite{sejl} $n_s=0.98\pm0.02$,
$\alpha=-0.003\pm0.010$. Combining the observations of three-year
WMAP, SDSS, SNIa and galaxy clustering \cite{se}, one can give the
constraints $n_s=0.965\pm0.012$, $\alpha=-(2.0\pm1.2)\times
10^{-2}$ and $r<0.22$.
These bounds of $n_s$ and $\alpha$ are all
at the $68\%$  confidence level, and that of $r$ is at $95\%$
 confidence level.
 These various bounds are consistent with each other,
taking into account of the corresponding confidence levels.
 Here in our calculation we
will choose the most loose constraints
 \be\label{33}
 n_s\in[0.86,~1.00]~,~~~\alpha\in[-0.087,~0.007]~,~~~
 r<0.22 ~~.
 \ee
which imply that  the primordial scalar spectrum is ``red" or
scale-invariant, and the running of scalar index is very small,
as required by the slow-roll
inflationary models.

The constraints on $\Omega_{g}$ mainly come from the
observations at small scale. Here we call them as ``small-scale
constraints" (SSC), which include the tightest constraint from the
observations of the pulsar timing \cite{timing}
 \be\label{35}
 \Omega_{g}h^2<2\times 10^{-9}, ~~~\nu=1.9\times10^{-9}Hz~;
 \ee
the constraint from the recent observations of LIGO \cite{LIGO},
 \be\label{36}
 \Omega_{g}<8.4\times 10^{-4},~~~69Hz<\nu<156Hz;
 \ee
and the constraint from the observations of BBN \cite{BBN,review},
 \be\label{37}
 w_{g}h^2<8.9\times10^{-6}~,
 \ee
where $w_{g}\equiv\int \Omega_{g}(\nu)d\ln \nu$.
Comparing with the constraints (\ref{limit0}) and (\ref{limit1}),
it is fair to say that the current SSC are too
loose to give any constraint on the single-field inflationary
models. This result will also be checked in the following
numerical calculation.


~\\
\emph{\textbf{The distribution of the realizations}}

In this part, we will present a numerical program
to solve the inflationary flow
equations (\ref{10})-(\ref{12}) for a large number of models,
where each initial condition randomly chosen
within  the constraints of (\ref{13})-(\ref{16})
will represent  an inflationary  model.

First, we want to study how tight
the LSC of Eq.(\ref{33}) and SSC of Eqs.(\ref{35})-(\ref{37}) are, as constraints,
on the inflationary models.
We have produced $10^7$ realizations of inflationary models.
It turns out that all these realizations satisfy the SSC, which
attests to the conclusion before: the current SSC are too loose to
give any actual constraint on the single-field inflationary
models. On the other hand, among these  $10^7$ realizations, only $5523$
of them ($\sim0.05\%$) satisfy the LSC of Eq.(\ref{33}).
So this
constraint is  tighter  for the inflationary models. In the
following, we will mainly discuss the distribution of these $5523$
realizations.

During the numerical calculations, the inflation can end in one of
the following two ways.
One is that $\epsilon<1$ is violated in the process of computing,
then the inflation automatically  stopped.
A number of of inflationary models are of this class,
such as the polynomial ``large-field" models, the
``small-field" polynomial potentials \cite{cata}.
The other way is by an abrupt termination,
perhaps from intervention of an auxiliary field as in hybrid inflation.
The linear potentials and
the exponential potentials also belong to this class \cite{cata}.
Here we choose the abrupt stop to be at $N=70$ in computation.
We have found that among these
$5523$ realizations, only $14$ of them stop the inflation in the
first way, all others do in the second way.
This fact is consistent
with the previous works \cite{attractor,before-work}. In following
we will discuss these two kinds of realizations separately.

First, we discuss the $5509$ realizations that exit inflation by abrupt termination.
They also satisfy  both the
large and small scale
constraints in Eqs.(\ref{33})-(\ref{37}).
These models and can inflate at least
$70$ e-folds. In Fig.[2], we plot them in the $r-\Omega_g$ plane,
which shows the following characters:

\emph{a}. For a fixed $r$, the distribution of $\Omega_g$ is very
scattered, especially at the region with large $r$.
For example, for a fixed $r=0.22$,
the values of $\Omega_g$ are distributed in a broad
range $\Omega_g\in[10^{-45},~10^{-20}]$;

\emph{b}. For each  fixed $r$, the values of $\Omega_g$ have an upper
limit,
and the small region just below this limit tend to contain most of the realizations;

\emph{c}. For each  fixed $r$,
 the values of the upper limit  $\Omega_g$ from our numerical result  are smaller
than the analytic results of Eq.(\ref{ome1}),
especially in the region $r>0.01$;

\emph{d}. At $r\simeq0.03$, the strength of RGW attains  the
maximum  value $\Omega_g\simeq2\times10^{-17}$,
which is almost an order of magnitude
smaller than the analytic result of Eq.(\ref{limit0}).
And this is beyond the sensitive ranges of LIGO, LISA, ASTROD and BBO.

\emph{e}. Most of the realizations tend to concentrate in the region with
the large values of $r$,
and the larger  $r$ is, the denser the distribution of the realizations are.
More than $90\%$ of realizations are in  the region of $r>0.01$.
This phenomenon of distribution may be due to
our specific choice of the initial conditions in Eq.(\ref{13})-(\ref{16}).

Among these $5509$ realizations, $50.21\%$ fall into the sensitivity
region of the Planck satellite,
$97.11\%$ fall into that of Clover,
$99.29\%$ fall into that of CMBPol,
and $42.91\%$ fall into the sensitivity region of the  DECIGO.
In comparison with the CMB observations,
much less realizations are in the sensitivity regions
of laser interferometer detectors. But DECIGO can detect the RGW with $r$
being much smaller than $10^{-4}$,
which is  beyond the CMB experiments.
This conclusion is the same as the analytic results in section 3.
Therefore, the CMB experiments and the laser interferometers are
complementary to each other for the RGW detection.

Now let us look at the $14$ realizations that satisfy all the
constraints in Eqs.(\ref{33})-(\ref{37}), but end the inflation
before the e-folds $N=70$ arrived. We found, for these
realizations, the values of e-folds are all in the region of
$N\in[40,~70]$, which is consistent with the current observations
and the theoretic predictions \cite{cata}. In Fig.[3], we plot
them in the $r-\Omega_g$ plane. This figure shows an interesting
feature: larger $r$ corresponds to a smaller $\Omega_g$, which is
also consistent with the distribution of realizations in Fig.[2].
Among these realizations, $35.71\%$  fall into the sensitive
region of the Planck satellite,  $100\%$ fall into the sensitive
regions of Clover and CMBPol, and $64.29\%$ fall into the
sensitive region of the  DECIGO. These results are also
consistent with the distribution of the  $5509$
realizations discussed in the above.

It should be mentioned that, in our numerical calculation,
 the initial conditions have been chosen randomly for the Hubble slow-roll
parameters in the regions (\ref{13})-(\ref{16}).
It is not clear which one is closer to the actual situation
of the inflationary process in the early Universe.
However, given the very broad range of initial conditions
for the Hubble slow-roll parameters,
these large sample of $10^7$ realizations
may exhaust, to some extent,
the reasonable reservoir  of inflationary models
driven by the single scalar field.

\begin{center}
{\large 5. Conclusion}
\end{center}
The relic gravitational waves is regarded as a strong
evidence for the inflationary models,
which is directly related to
the energy scale of the inflation.
Although up to  now people
have not  observed RGW, a lot of constraints
have been obtained on them.
 These constraints include two kinds: One is the LSC, which
are mainly from the CMB observations, especially the recent WMAP
results.
This can constrain the RGW at very low frequency
$\nu\in[10^{-17},~10^{-15}]$Hz.
The other is the SSC, mainly from LIGO, BBN, and pulsar timing,
and it is sensitive to the waves with high frequencies.
A number of experiments
are under development for the RGW detection,
which can be classified as the following types:
the CMB experiments
(sensitive to the waves with $\nu\in[10^{-17},~10^{-15}]$Hz),
including Planck, Clover, CMBPol, etc,
the laser interferometers
(sensitive to the waves with $\nu\in[10^{-4},~10^{4}]$Hz),
including advanced
LIGO, LISA, ASTROD, BBO, DECIGO, etc,
the electromagnetic detectors
 (sensitive to the waves with $\nu\in[10^{8},~10^{10}]$Hz).

We have calculated the strength of RGW,
studied how tight of the current constraints on the RGW,
and examined the detective abilities of the future experiments.
When calculating the values of $\Omega_g(k)$,
we have worked in  two methods:
the analytic and the numerical.
The former method simply shows the
dependent relation of $\Omega_g(k)$ on the inflationary
parameters $n_s$ and $r$.
After considering the current
constraints on these parameters,
we have given  an upper limit
$\Omega_g<3.89\times10^{-16}$,
where the three damping effects have been included,
such as the reashift-suppression, the accelerating expansion,
and the free-streaming neutrinos.
This limit is in the
sensitivity  ranges of BBO and DECIGO, but beyond which of advanced
LIGO, LISA and ASTROD.
In the numerical method,
the energy spectrum $\Omega_g(k)$ has been calculated
by solving the inflationary flow equations.
The resulting spectrum is more precise for
the RGW at high frequency range.
The corresponding  upper limit $\Omega_g<8.62\times10^{-14}$
has been given,
which is independent of the inflationary parameters and applies to
any single-field slow-roll inflationary model.
After considering
the constraints on $n_s$, $\alpha$ and $r$, this bound becomes
$\Omega_g<2\times10^{-17}$, which is beyond the sensitivity  limit of
BBO.

The results from these two methods suggest  the consistent
conclusions: The current constraints on the RGW from LIGO, BBN and
pulsar timing are too loose to give any constraint for the
single-field inflationary models, and the constraints from CMB and
LSS are relatively tighter.
The future laser interferometer,
DECIGO, is more effective for detecting the RGW with smaller $r$,
but the CMB experiments, such as Planck, Clover and CMBPol, are more
effective for detecting waves with larger $r$.
They are complementary to each other for the RGW detection.
The laser
interferometers, as the advanced LIGO, LISA and ASTROD have
less chances to find the signal of RGW, if the single-field
inflationary model is held.

A final remark should be made,
that is, all conclusion on RGW and their detection constraints
arrived in this paper
are pertinent only for single scalar field  models for inflation.
The RGW generated from other models of inflation need to
be analyzed separately.

~

\textbf{ACKNOWLEDGMENT}: We thank S.Chongchitnan for helpful
discussion. W.Zhao has been partially supported by Graduate
Student Research Funding from USTC. Y.Zhang has
been supported by the Chinese NSF (10173008), NKBRSF (G19990754),
and by SRFDP.


\newpage

\begin{figure}
\centerline{\includegraphics[width=15cm]{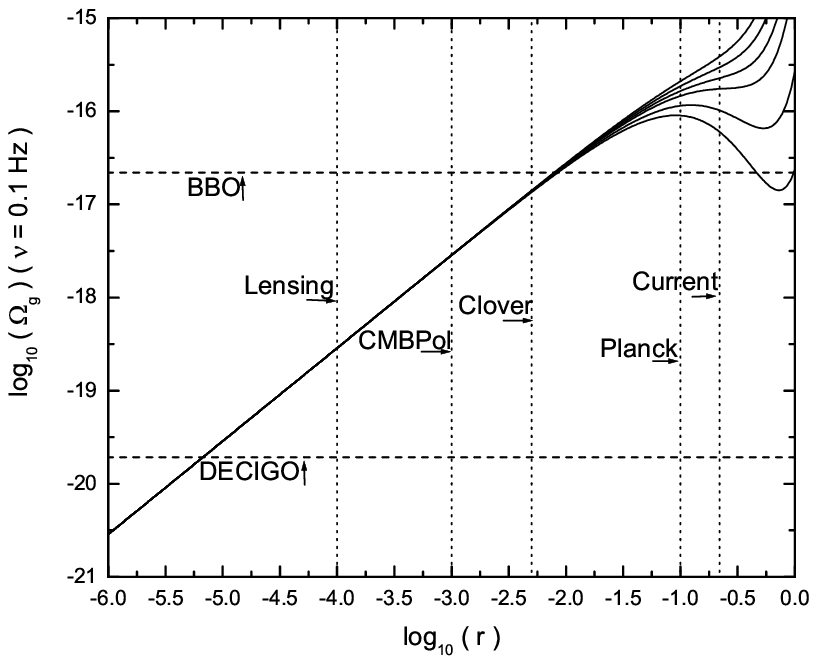}}
 \caption{\small The strength of RGW at $\nu=0.1$Hz depends
 on the slow-roll parameters $n_s$ and $r$.
 This figure shows the results of
 analytic approximation in (\ref{ome1}).
 The solid lines from top down are the curves with
 $n_s=1.00,0.98,0.96,0.94,0.90,0.86$, respectively.
 The vertical (dot) lines from right to
 left are the sensitive limit curves of current observations,
 Planck, Clover,
 CMBPol, and the sensitive limit of CMB observations, respectively.
 The horizontal (dash)
 lines from up to down are the sensitive limit curves of BBO
 and  DECIGO, respectively. }
\end{figure}

\begin{figure}
\centerline{\includegraphics[width=15cm]{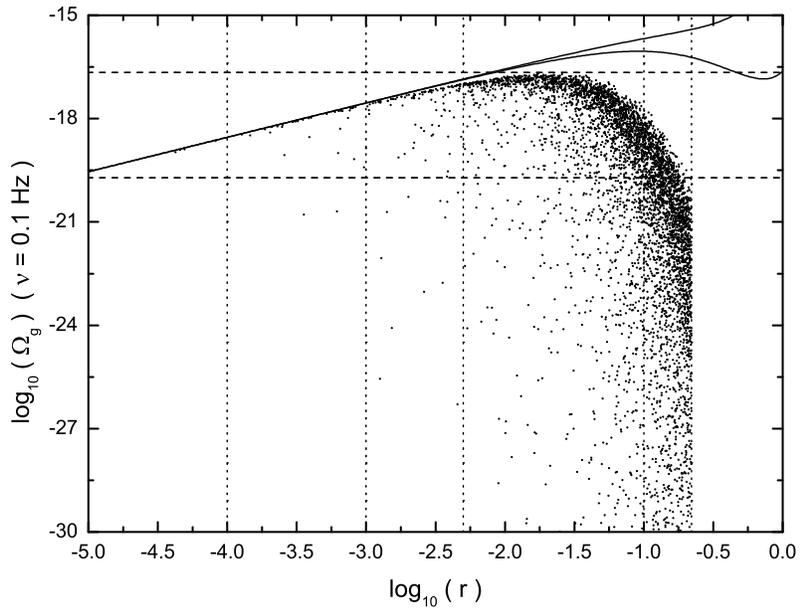}}
 \caption{\small The distribution of the $5509$
 realizations in the $r-\Omega_g$ plane.
 The solid lines from top down are the
 analytic curves with $n_s=1.00$ and
 $n_s=0.86$, respectively. The vertical (dot)
 lines and the horizontal (dash)
 lines have the same meanings with which in Fig.[1].  }
\end{figure}

\begin{figure}
\centerline{\includegraphics[width=15cm]{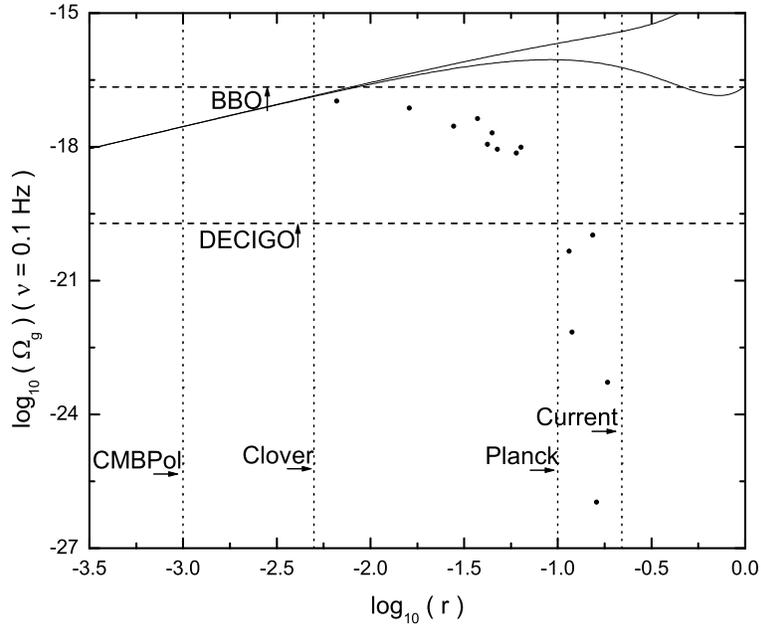}}
 \caption{\small The distribution of the $14$ realizations
 in the $r-\Omega_g$ plane.
  The solid lines from top down are the analytic
  curves with $n_s=1.00$ and
 $n_s=0.86$, respectively.
 The vertical (dot) lines from right to
 left are the sensitive limit curves of current observations,
 Planck, Clover and
 CMBPol, respectively. The horizontal (dash)
 lines from up to down are the sensitive limit
 curves of BBO and DECIGO, respectively. }
\end{figure}

\end{document}